\documentclass{ws-procs9x6}
\begin{document}
\title{Adaptive spectral identification techniques in presence of undetected
  non linearities}
\author{G. CELLA}
\address{Universit\'a di Pisa and I.N.F.N., \\
Via Buonarroti 2, 56100 Pisa, Italy\\ 
E-mail: giancarlo.cella@pi.infn.it}

\author{E. CUOCO}
\address{INFN Firenze, \\ 
Via G. Sansone 1,50019 Sesto F. (Firenze), Italy\\
E-mail: cuoco@fi.infn.it}  

\author{G.~M. GUIDI}
\address{Istituto di Fisica, Universit\'a degli Studi di Urbino, \\ 
Via S. Chiara 27, 61029 Urbino, Italy\\
E-mail: guidi@fis.uniurb.it}




\maketitle

\abstracts{The standard procedure for detection of gravitational wave
 coalescing binaries signals is based on Wiener filtering with an
 appropriate bank of template filters. This is the optimal procedure
 in the hypothesis of addictive Gaussian and stationary noise. We
 study the possibility of improving the detection efficiency with a
 class of adaptive spectral identification techniques, analyzing their
 effect in presence of non stationarities and undetected non
 linearities in the noise.}

\section{Adaptive whitening filter and non stationarities}
We can define the output of an interferometric detector $x(t)$ as the
sum of noise $n(t)$ and signal $h(t,\mathbf{\theta})$ of known shape
and unknown parameters $\mathbf{\theta}$:
\begin{equation}
x(t) = n(t) + h(t,\mathbf{\theta}).
\end{equation}
It is well established that the optimal filter for the detection of
the signal $h(t,\theta)$, if the noise is Gaussian and stationary, is
the Wiener matched filter\cite{fi}:
\begin{equation}
C(\mathbf{\theta}) = \int \frac{\tilde{x}(\nu) \tilde{h}^{*}(\nu,\mathbf{\theta})}{S(\nu)} d\nu
\end{equation}
\noindent being $S(\nu)$ the noise Power Spectral density (PSD) and $h(\nu,\mathbf{\theta})$ the
template for the signal we are looking for.

\begin{figure}[ht]
\centerline{\psfig{file=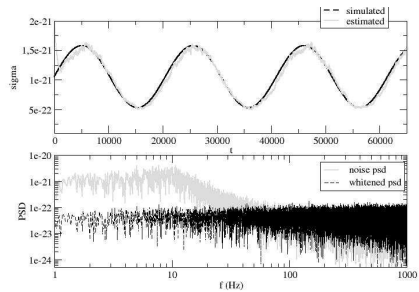,width=.78\textwidth,bbllx=0pt,bblly=50pt,bburx=792pt,bbury=560pt,clip=}}   
\caption{Top plot: standard deviation $\sigma$ of the non-stationary
data -$A= 0.5$ and $\nu= 0.1$- and adaptive  estimated standard
deviation: the LSL filter is able to follow the non stationarity with $\lambda=0.997$. Bottom plot: PSD of the non stationary
data before and after whitening.\label{inter}}
\end{figure}

If the noise is not stationary we have problems in estimating the
PSD. We can avoid these by generalizing in time domain the Fourier
space the "whitening" procedure in the Wiener filter, using an adaptive
filter: with the "whitening" we transform a colored data sequence in a
white one, i.e. $\delta$-correlated with a "flat" PSD. We will use a
linear filter produced by fitting the PSD with and Auto Regressive
(AR) model\cite{cu1,cu2,al}:
\begin{equation}
\label{eq:xn}
x[n] = \sum_{k=1}^{P} a_{k} x[n-k] + \sigma w[n]
\end{equation}
where $a_{k}$,$\sigma$ are the parameters of the model, $P$ the order of
the model and $w[n]$ a driving normal white noise. The idea of
adaptive filters is to find the optimal weights $a_{k}$ which minimize
a cost function error $\epsilon$. For Least Squares based methods this is
\begin{equation}
\label{eq:en}
\epsilon(n) = \sum_{i=1}^{n} \lambda^{n-i} e^{2}[i|n]
\end{equation}

\noindent where $e[i|n]$ is the error done in the prediction at time $n$ with the
$a_{k}$ calculated at time $n$. The parameter $\lambda$, where $0<
\lambda \leq 1$, is called "forgetting factor" and is introduced to
let the filter follow the changes in the parameters of the noise
model~\cite{al}.  The adaptive filter we used, obtained by minimizing eq.~\ref{eq:en}, is the Least Squares Lattice (LSL) filter~\cite{al}. When we adaptively identified the parameters and their time variation with the LSL filter, we can use this filter to whiten the data: $x[n]$ becomes the input colored sequence
which enters in the filter, and $w[n]$ is the output whitened sequence. We simulated a Virgo-like~\cite{Virgo} non stationary noise by fitting
the $a_{k}$,$\sigma$ parameters on the theoretical PSD for thermal and
shot noise of the Virgo interferometer~\cite{cu1} and adding a
sinusoidal modulation to $\sigma$:
\begin{equation}
\sigma[n] = \sigma (1+A \sin 2 \pi \nu n)
\end{equation}
With the forgetting factor $\lambda$ less than one 
we succeeded in following the non-stationarity and in whitening the
non stationary process obtaining a flat PSD as it is expected for a
white process. In the first plot of Fig.~\ref{inter} we compare the
standard deviation of the simulated non stationary time series with
the standard deviation estimated by  LSL filter, whereas in
the second plot the PSD of the simulated non stationary time series
and the PSD of the whitened time series  are shown. The use of the adaptive filter with $\lambda
< 1$ is important in view of the occurrence of unknown slow non
stationarity in the interferometer data, where we cannot apply anymore
the optimal filter.

\section{Performance of whitening and non-linearities}

Apparent non stationarities can appear as the effect of a stationary
non linear process. For example, if we observe a spike, this can be a
signature of both non stationarity or non linearity, whereas a linear
stationary process has a low probability to be ``spiky''. The
whitening filter should then show a good performance also in presence
of non linearities, because we may not be able to know which type of
process we are analyzing.

We simulated a non linear process for a Virgo-like model with an AR
model with conditional heteroscedasticity (ARCH)\cite{ha}. 
\begin{figure}[ht]
\centerline{\psfig{file=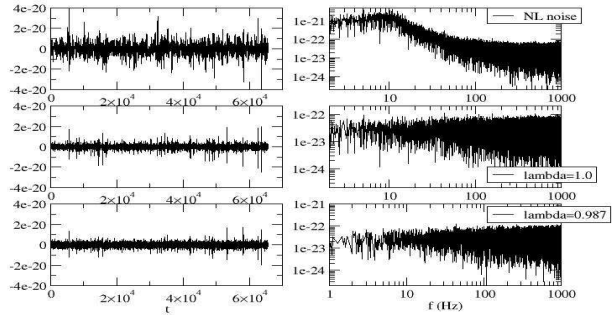,width=\textwidth,height=180pt,bbllx=65pt,bblly=50pt,bburx=765pt,bbury=552pt,clip=}}   
\caption{Performance of the adaptive filter in presence of non linearities. The plots on the right are the time
  series obtained with: the ARCH model, the same model filtered with
  $\lambda=1$ and with $\lambda=0.987$. The plots on the right are the
  PSD in the three different cases. \label{inter2}}
\end{figure}
The time evolution of a Virgo-like ARCH model is reproduced in
Fig.~\ref{inter2}, first plot: it presents a lot of spikes, but its
PSD (on the right) is almost the same of the linear Virgo-like AR
model, as the PSD retains only the linear features of a process.  The
whitening filter with $\lambda=1$, when applied to this process (plots
on the second line of Fig.~\ref{inter2}) does not completely eliminate
the spikes, as the non-linearities are averaged away in the limit of
an infinite number of data and are therefore irrelevant in the
spectral estimation: the PSD is flat and the process has successfully
been whitened, but the spikes could give rise to false alarms in the
detection process, being identified as signals. For $\lambda=0.987$
(Fig.~\ref{inter2}, third line) we find a reduction of the spikes. This
can help in reducing the false alarm probability.

\section{Conclusions}
We showed the important role of forgetting factor in the adaptive LSL filter in identifying non stationarities of time series processes.
Moreover this adaptive whitening procedure is apparently effective in reducing
both non stationarity and non linearity effects. This could be
important in order to improve the robustness of algorithm for
gravitational wave detection~\cite{vi3}.

\end{document}